\documentclass[conference]{IEEEtran}
\IEEEoverridecommandlockouts
\usepackage{cite}
\usepackage{amsmath,amssymb,amsfonts}
\usepackage{algorithmic}
\usepackage{tikz}
\usepackage{pgfplots}
\usepackage{graphicx}
\usepackage{subcaption}
\usepackage{multirow}
\usepackage{textcomp}
\usepackage{xcolor}
\usepackage{soul}
\usepackage{fancyhdr}

\fancypagestyle{icecsheader}{
    \fancyhf{}
    \fancyhead[C]{%
        \fontsize{8}{9}\selectfont
        2026 IEEE 33rd International Conference on Electronics, Circuits and Systems (ICECS)}

}

\fancypagestyle{icecsfirstpage}{
    \fancyhf{}
    \fancyhead[C]{%
        \fontsize{8}{9}\selectfont
        2026 IEEE 33rd International Conference on Electronics, Circuits and Systems (ICECS)}
    \fancyfoot[L]{%
        \fontsize{8}{9}\selectfont
        979-8-3195-1905-4/26/\$31.00 \textcopyright{}2026 IEEE}

}

\renewcommand{\baselinestretch}{0.934}

\newcommand\freefootnote[1]{%
  \let\thefootnote\relax%
  \footnotetext{#1}%
  \let\thefootnote\svthefootnote%
}

\def\BibTeX{{\rm B\kern-.05em{\sc i\kern-.025em b}\kern-.08em
    T\kern-.1667em\lower.7ex\hbox{E}\kern-.125emX}}
\begin{document}

\title{Reinforcement Learning-Based Optimization of Workload-Aware Power Delivery Networks

}

\author{
    \IEEEauthorblockN{Oran Hayes\textsuperscript{†}, Maria Pantazi-Kypraiou\textsuperscript{\ddag}, Athanasios Tziouvaras\textsuperscript{\ddag},~George Stamoulis\textsuperscript{\ddag},~Anuj Pathania\textsuperscript{¶},\\ Shreejith Shanker\textsuperscript{†},~and~George Floros\textsuperscript{†}}
    \IEEEauthorblockA{
    \textsuperscript{†}Trinity College Dublin, Ireland, emails: \{hayesor, shankers, florosg\}@tcd.ie,  \\
    \textsuperscript{\ddag}University of Thessaly, Greece, emails: \{mpantazi-, attziouv, georges\}@e-ce.uth.gr \\
\textsuperscript{¶}University of Amsterdam, The Netherlands, email: a.pathania@uva.nl\\
}}

\maketitle

\begin{abstract}
Power Delivery Networks (PDNs) are critical components of modern VLSI chips, providing stable voltage levels while satisfying electromigration (EM) and IR-drop constraints. Conventional PDN design methodologies typically rely on worst-case assumptions, often resulting in over-provisioned networks and inefficient use of  resources. This paper presents a reinforcement learning-based framework for the optimization of workload-aware PDNs. The proposed methodology first generates workload-aware PDNs using architectural power traces obtained from system-level simulations. These power traces are mapped to spatial power density distributions, enabling adaptive allocation of PDN resources according to local current demand. A reinforcement learning agent then performs wire-width optimization to minimize PDN area while maintaining EM and voltage integrity constraints. Electrical and reliability metrics are obtained using SPICE-based circuit analysis and EM lifetime estimation. Experimental evaluation is performed on a dataset of workload-aware PDNs generated from 4-, 8-, and 16-core multiprocessor floorplans using PARSEC and SPLASH-2 benchmark workloads. Furthermore, the proposed  Deep Q-Network (DQN)-based optimizer reduces the average normalized PDN area by 47\% while satisfying all EM and IR-drop constraints. Compared to simulated annealing, the proposed approach achieves comparable optimization quality while providing approximately 26$\times$ faster optimization.
\end{abstract}


 
\section{Introduction}
The increasing integration density of modern multicore systems has significantly intensified the challenges associated with on-chip power delivery \cite{9377004}. \freefootnote{*This work was funded under the COIN-3D project, which has received funding from the European Union’s Horizon Europe research and innovation program under grant agreement No. 101159667.}Emerging technologies such as backside power delivery networks (BS-PDNs) further highlight the importance of efficient power distribution by enabling improved routing flexibility, reduced IR drop, and better utilization of routing resources \cite{sisto2023block}. However, regardless of the underlying PDN architecture, designers must still balance competing objectives including metal area, power integrity, and reliability. Traditional PDN design methodologies are typically based on conservative worst-case assumptions, often resulting in over-provisioned networks that consume excessive routing resources and chip area \cite{hayes2026workload}. At the same time, electromigration (EM) and IR-drop constraints limit the ability to aggressively reduce PDN resources, creating a complex optimization problem for modern VLSI systems.

To this end, several methodologies have been proposed for PDN optimization under area, reliability, and voltage integrity constraints. Early approaches relied on linear and nonlinear programming techniques to minimize routing resources while satisfying EM/IR requirements \cite{tan1999reliability}. More recently, physics-based EM models have enabled more accurate reliability-aware optimization \cite{moudallal2019power}, leading to power-grid repair and wire-sizing methodologies based on mathematical optimization, metaheuristics such as Particle Swarm Optimization (PSO) \cite{glsvlsi}, and gradient-based approaches including L-BFGS \cite{axelou2025towards}. In parallel, workload-aware PDN design methodologies have emerged, exploiting architectural power traces to allocate resources according to realistic current demand distributions rather than worst-case assumptions. Despite these advances, existing approaches largely depend on predefined optimization strategies and repeated design-space exploration. The use of reinforcement learning for workload-aware PDN optimization remains largely unexplored, particularly for learning adaptive wire-sizing policies that minimize routing resources while satisfying EM and IR-drop constraints.

In this paper, we present a reinforcement learning-based framework for the optimization of workload-aware PDNs. The main contributions of this work are threefold. \textit{First}, we formulate workload-aware PDN optimization as a reinforcement learning problem, enabling an agent to automatically learn wire-sizing strategies from interaction with the design environment. \textit{Second}, we integrate workload-dependent power distributions derived from architectural simulations into the optimization process, allowing PDN resources to be allocated according to realistic current demands rather than worst-case assumptions. Finally, we evaluate the proposed approach on workload-aware PDNs generated from multicore processor floorplans using representative benchmark applications and compare its performance against simulated annealing. Experimental results show that the proposed DQN-based optimizer achieves near-optimal solutions while requiring significantly lower computational cost than a simulated annealing approach.

The remainder of this paper is organized as follows. Section~\ref{background} presents the background for PDN design. Section~\ref{proposed} describes the proposed reinforcement learning-based optimization framework. Experimental setup and results are presented in Section~\ref{exp}, followed by conclusions in Section~\ref{cocl}.

\section{Background}\label{background}
\subsection{Workload-Aware PDN Design}

PDNs are traditionally designed using uniform or worst-case power assumptions to guarantee reliable operation under all possible operating conditions. In this approach, routing resources are allocated conservatively across the entire chip, resulting in regular power grid structures that are often over-provisioned \cite{4483978}. While such designs provide sufficient voltage integrity and EM robustness, they may consume excessive metal resources and reduce the routing capacity available for signal interconnects.

In modern multicore systems, power consumption is highly dependent on workload behavior and varies significantly across architectural components. Workload-aware PDN design seeks to exploit this information by incorporating power estimates obtained from architectural simulations into the PDN planning process. Given a supply voltage $V_{DD}$, the current demand of a region can be estimated as:

\begin{equation}
I_i = \frac{P_i}{V_{DD}},
\label{eq:current_demand}
\end{equation}
where $P_i$ denotes the power consumed within region $i$. If the region occupies an area $A_i$, the corresponding power density can be expressed as

\begin{equation}
\rho_i = \frac{P_i}{A_i},
\label{eq:power_density}
\end{equation}
where $\rho_i$ represents the spatial power density. These metrics provide a representation of workload-dependent current demand across the floorplan and can be used to identify regions with different power requirements.

Based on the estimated power distribution, PDN resources can be allocated adaptively according to local current demand \cite{hayes2026workload}. Regions with higher power density are assigned denser power grids to support larger currents, while regions with lower activity require fewer routing resources. Compared to conventional worst-case methodologies, workload-aware PDNs can reduce overall metal usage while maintaining voltage integrity and reliability constraints, providing a more efficient starting point for subsequent optimization.
\subsection{Electromigration in Power Delivery Networks} 

EM is a major reliability concern in modern PDNs. It is caused by the momentum transfer from conducting electrons to metal atoms, leading to gradual material transport within interconnects. Over time, this process can result in void formation, increased resistance, and ultimately open-circuit failures. As technology nodes continue to scale and current densities increase, accurate EM assessment has become an important consideration in PDN optimization.

One of the earliest and most widely used EM lifetime models is Black's equation \cite{black2005electromigration}, which estimates the mean time to failure (MTTF) of a conductor as:

\begin{equation}
\mathrm{MTTF} = A J^{-n} e^{\frac{E_a}{kT}}
\end{equation}

where $J$ is the current density, $T$ is the operating temperature, $E_a$ is the activation energy, $k$ is Boltzmann's constant, and $A$ and $n$ are empirical fitting parameters. Black's equation provides a simple relationship between current density and expected lifetime, making it attractive for fast reliability estimation. However, it treats each interconnect segment independently and does not account for stress interactions between neighboring wires in a power grid.

To address these limitations, a physics-based stress evolution model was proposed in \cite{korhonen1993stress}, which describes the temporal evolution of hydrostatic stress within a conductor:

\begin{equation}
\frac{\partial \sigma}{\partial t}
=
\frac{B\Omega}{kT}
\frac{\partial}{\partial x}
\left[
D_a
\left(
\frac{\partial \sigma}{\partial x}
-
\frac{q^{*}\rho}{\Omega}J
\right)
\right]
\end{equation}

where $\sigma$ denotes hydrostatic stress, $D_a$ is the atomic diffusivity, $B$ is the bulk modulus, $\Omega$ is the atomic volume, $\rho$ is the resistivity, and $q^{*}$ is the effective charge number. Unlike Black's equation, Korhonen's model captures the spatial redistribution of stress and interactions between connected interconnect segments, providing a more realistic representation of EM degradation in mesh-like PDNs. The increased accuracy comes at the cost of significantly higher computational complexity, making direct use of the model challenging within iterative optimization frameworks.

\begin{figure*}[!hbt]
    \begin{center}
        \includegraphics[width=1.95\columnwidth]{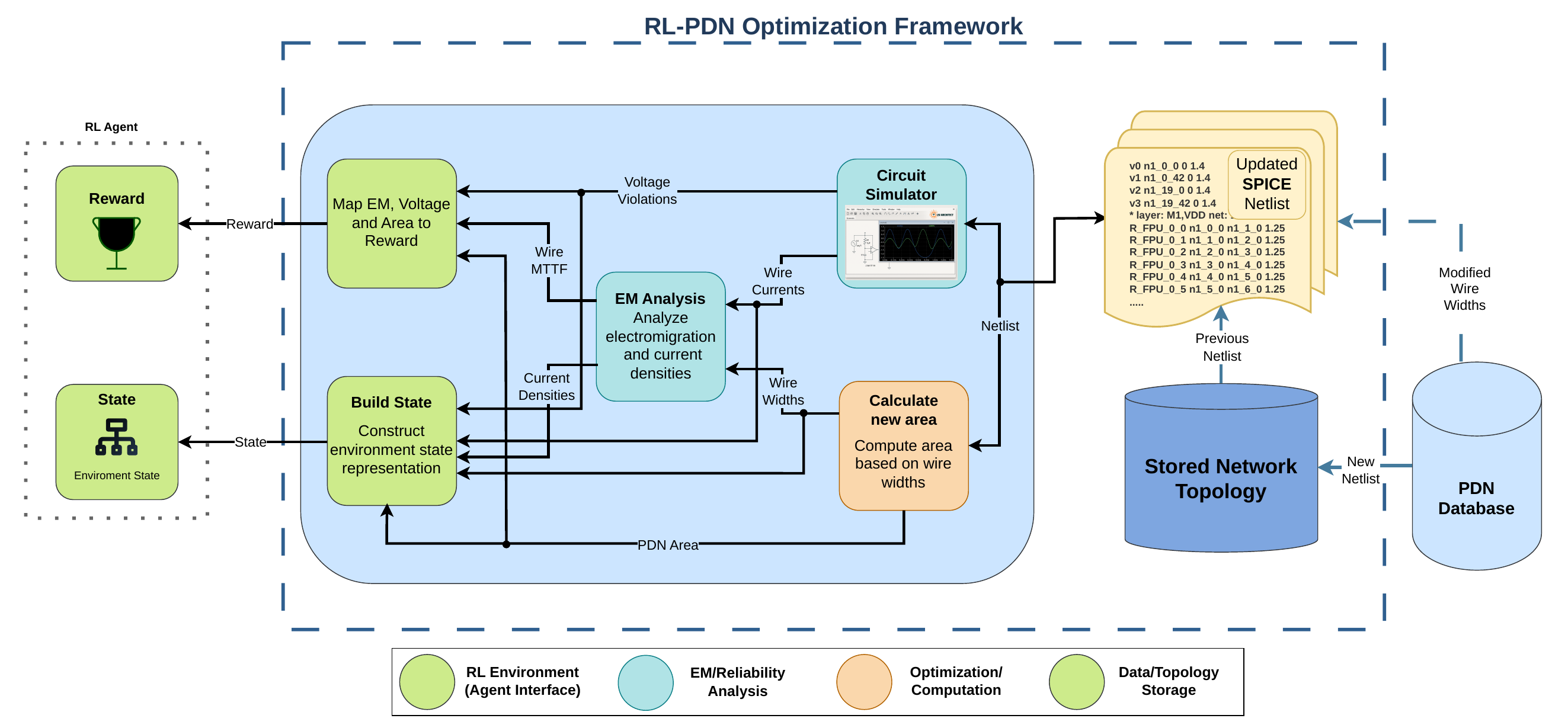}
        \caption{Interaction between the RL agent and the PDN optimization environment.}
\label{fig:RL_meth}
    \end{center}
\end{figure*}

\section{Reinforcement Learning-Based Optimization of Workload-Aware PDNs}\label{proposed}

\subsection{Problem Formulation and EM-Constrained Environment}

The objective of the optimization process is to minimize the routing area of a  PDN while satisfying EM and voltage integrity constraints. Given a fixed PDN topology, optimization is performed by adjusting the widths of individual wires. Let $w_i$ and $L_i$ denote the width and length of wire $i$, respectively. The optimization problem can be expressed as:

\begin{equation}
\min \sum_{i=1}^{N} w_iL_i
\label{eq:objective}
\end{equation}
subject to wire-width limits, EM constraints, and allowable voltage-drop bounds \cite{glsvlsi}.

A key challenge in reinforcement learning-based optimization is the repeated evaluation of candidate solutions. During training, thousands of PDN configurations may be explored, requiring fast estimation of both electrical and reliability metrics. While the Korhonen stress-evolution model provides an accurate representation of EM behavior, solving the corresponding partial differential equations for every RL step would introduce a prohibitive computational overhead.

To enable efficient optimization, a surrogate EM model based on Black's equation is employed during training. Black's equation estimates the MTTF of a conductor as:

\begin{equation}
\mathrm{MTTF}=A'J^{-n},
\label{eq:black}
\end{equation}
where $J$ denotes the current density and $A'$ and $n$ are technology-dependent fitting parameters. The parameters of Black's model are calibrated using time-to-failure data generated from the physics-based Korhonen model. Taking the logarithm of (\ref{eq:black}) yields:

\begin{equation}
\log(\mathrm{MTTF})=-n\log(J)+\log(A'),
\label{eq:black_log}
\end{equation}
which enables the parameters to be obtained through linear regression.

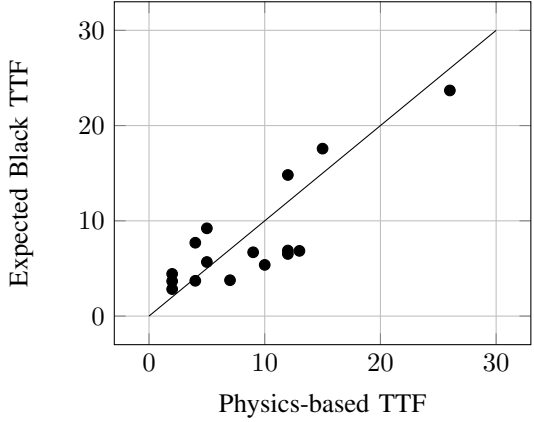
\begin{figure}[h]
    \centering
    \begin{tikzpicture}
    \begin{axis}[
        width=0.8\linewidth,
        xlabel={Physics-based TTF},
        ylabel={Expected Black TTF},
        grid=both,
    ]

    \addplot[
        only marks,
    ]
    coordinates {
        (2,4.43008678)
        (5,9.214396247)
        (12,14.81477859)
        (26,23.68977927)
        (15,17.5800229)
        (4,7.700746191)
        (2,3.664022507)
        (5,5.678376149)
        (9,6.698860922)
        (12,6.882551549)
        (13,6.846828355)
        (12,6.522729168)
        (10,5.375995629)
        (7,3.767071125)
        (4,3.70472751)
        (2,2.833245271)
    };

    \addplot[
        domain=0:30,
    ]
    {x};

    \end{axis}
    \end{tikzpicture}
    \caption{Comparison of physics-based TTF and Expected TTF from Black’s equation.}
    \label{fig:blacks_calibration}
\end{figure}

Figure~\ref{fig:blacks_calibration} compares the TTF predicted by the calibrated Black model against the corresponding values obtained from physics-based simulations. The observed agreement demonstrates that the calibrated model provides a sufficiently accurate approximation for use within the optimisation loop while reducing computational cost by several orders of magnitude.

\subsection{RL Optimization Framework}
Figure~\ref{fig:RL_meth} depicts the proposed optimization environment. At the beginning of each episode, a workload-aware PDN is loaded from the dataset and its SPICE netlist is parsed to extract the network topology and wire-width information. The topology remains fixed throughout the optimization process while the wire widths are modified by the agent.

After an action is selected, the corresponding resistor values in the SPICE netlist are updated to reflect the new wire widths. The modified netlist is then analyzed using a circuit simulator to obtain node voltages and branch currents. These electrical quantities are used to evaluate the routing area, voltage-drop violations, and EM reliability of the current design.

To enable efficient optimization, EM is estimated using the calibrated Black model described previously. The resulting area, voltage, and EM metrics are combined to construct the environment state and reward signal, which are returned to the agent for the next optimization step. This process is repeated until a termination condition is reached.

\subsection{State, Action, and Reward Design}
The state representation captures both the physical structure and electrical behavior of the PDN. At each optimization step, the state consists of the wire widths, current densities, and node voltage drops obtained from circuit simulation:

\begin{equation}
s_t =
\left[
w_1,\ldots,w_N,
J_1,\ldots,J_N,
V_1,\ldots,V_M
\right].
\label{eq:state}
\end{equation}

The action space consists of discrete wire-sizing operations. For each wire, the agent may increase or decrease its width by a fixed increment $\Delta w$,

\begin{equation}
w_i \leftarrow w_i \pm \Delta w,
\label{eq:action}
\end{equation}
subject to technology-dependent minimum and maximum width constraints. A no-operation action is also included.

The reward function is designed to minimize PDN area while enforcing EM and voltage integrity requirements. Let $A_n$ denote the normalized PDN area. A wire is considered to violate the EM constraint if its current density exceeds the maximum allowable current density $J_{\max}$ obtained from the calibrated Black model. Similarly, a voltage violation occurs when a node exceeds the allowable voltage-drop limit. The reward is defined as

\begin{equation}
R_t = -A_n
-\lambda_{EM}N_{EM}
-\lambda_{IR}N_{IR},
\label{eq:reward}
\end{equation}

where $N_{EM}$ and $N_{IR}$ denote the number of EM and voltage-drop violations, respectively, and $\lambda_{EM}$ and $\lambda_{IR}$ are penalty coefficients. This formulation encourages the agent to reduce routing resources while maintaining a valid PDN design.

\subsection{Deep Q-Network}
Finally, a Deep Q-Network (DQN) is used to approximate the action-value function $Q(s,a)$ with a neural network. This avoids explicitly storing a Q-table for the exponentially large PDN state space and enables generalization across similar wire-sizing configurations.

The Q-network evaluates each wire using both local and global PDN information. For each wire, local features are concatenated with global features describing the overall PDN state. These inputs are processed by a shared multilayer perceptron consisting of two fully connected layers with 128 neurons and ReLU activations. The network then branches into two heads: a per-wire head that outputs Q-values for increasing or decreasing each wire width by $\pm \Delta w$, and a global head that outputs the Q-value of the no-operation action.

The network parameters are trained by minimizing the temporal-difference loss

\begin{equation}
\mathcal{L} =
\left(
r_t + \gamma \max_{a'} Q_{\mathrm{target}}(s_{t+1},a')
-
Q(s_t,a_t)
\right)^2 ,
\label{eq:dqn_loss}
\end{equation}
where $r_t$ is the reward, $\gamma$ is the discount factor, and $Q_{\mathrm{target}}$ is a target network used to stabilize training. Experience replay is used to reduce temporal correlation between updates, while an $\epsilon$-greedy policy balances exploration and exploitation during training. Once trained, the DQN agent rapidly predicts wire-sizing actions for new PDN states, reducing the need for repeated search-based optimization.

\section{Experimental Results} \label{exp}
The proposed reinforcement learning framework was evaluated on a dataset of  workload-aware PDNs generated from 4-, 8-, and 16-core floorplans using PARSEC and SPLASH-2 workloads. The dataset contains both regular and irregular grid topologies under average and worst-case power scenarios. The DQN agent was trained on the generated PDNs and evaluated on unseen test instances. Performance was compared against the original workload-aware PDNs (Baseline) and a Simulated Annealing (SA) optimizer using the same objective and constraint definitions.

\begin{table}[t]
\centering
\caption{Comparison of the baseline PDNs, DQN, and SA. Area values are normalized.}
\label{tab:dqn_sa_comparison}
\begin{tabular}{l|cccc}
\hline
PDN & Method & Area & EM/IR Viol. & Runtime (s) \\
\hline \hline
\multirow{3}{*}{4-core}
& Baseline & 0.3330 & 0 / 0 & -- \\
& DQN      & 0.1724 & 0 / 0 & 72.73 \\
& SA       & 0.1450 & 0 / 0 & 1964 \\
\hline
\multirow{3}{*}{8-core}
& Baseline & 0.3120 & 0 / 0 & -- \\
& DQN      & 0.1728 & 0 / 0 & 148.63 \\
& SA       & 0.1438 & 0 / 0 & 3894 \\
\hline
\multirow{3}{*}{16-core}
& Baseline & 0.3050 & 0 / 0 & -- \\
& DQN      & 0.1608 & 0 / 0 & 294.61 \\
& SA       & 0.1425 & 0 / 0 & 7827 \\
\hline
\multirow{3}{*}{Average}
& Baseline & 0.3170 & 0 / 0 & -- \\
& DQN      & 0.1686 & 0 / 0 & 171.99 \\
& SA       & 0.1438 & 0 / 0 & 4561.67 \\
\hline
\end{tabular}
\end{table}

Table~\ref{tab:dqn_sa_comparison} compares the proposed DQN approach against SA on the test set. Both methods satisfy all EM and voltage constraints, yielding reliability-compliant solutions. While SA achieves a slightly lower average normalized area (0.1438 versus 0.1686), the difference in optimization quality is relatively small compared to the substantial difference in runtime. DQN completes optimization in an average of 171.99s, whereas SA requires 4561.67s, corresponding to an approximately 26$\times$ speedup. Therefore, the proposed DQN approach offers a more favorable trade-off between optimization quality and computational efficiency, achieving near-optimal area reduction at a fraction of the optimization cost.

\section{Conclusions}\label{cocl}

This paper presented a reinforcement learning-based framework for the optimization of workload-aware PDNs. The proposed methodology combines workload-dependent power distributions with a DQN agent that performs reliability-aware wire-width optimization under EM and IR-drop constraints. Experimental results on 4-, 8-, and 16-core PDNs demonstrate that the proposed approach reduces the average  PDN area from 0.3170 to 0.1686, corresponding to a 47\% reduction, while maintaining zero EM and voltage violations. Compared to simulated annealing, the DQN agent
achieves comparable optimization quality while providing
approximately 26$\times$ faster optimization, reducing the average
runtime from 4561.67s to 171.99s. These results demonstrate that reinforcement learning offers a favorable trade-off between routing-area reduction and computational cost for PDN optimization.

\bibliographystyle{IEEEtran}
\bibliography{ref}
\end{document}